# Neutrino Mixing and Future Solar Neutrino Experiments[★]

S.M. Bilenky[(a,b,c)] and C. Giunti[(b,c)]

[(a)] Joint Institute for Nuclear Research, Dubna, Russia
[(b)] INFN Torino, Via P. Giuria 1, I–10125 Torino, Italy
[(c)] Dipartimento di Fisica Teorica, Università di Torino

Possibilities of a model independent treatment of the data from future real-time solar neutrino experiments (SNO, Super-Kamiokande and others) are discussed. It is shown that in the general case of transitions of the initial solar $\nu_e$'s into $\nu_\mu$ and/or $\nu_\tau$ the total flux of initial $^8$B neutrinos and the $\nu_e$ survival probability can be determined directly from the experimental data. Lower bounds for the probability of transition of solar $\nu_e$'s into all possible sterile states are derived and expressed through measurable quantities.
★ Talk presented by S.M. Bilenky at TAUP93, LNGS (Assergi, Italy), September 1993. DFTT 66/93.

hep-ph/9310352  26 Oct 1993

## 1. Introduction

As is well known [1], the existing solar neutrino data can be described by the resonant MSW enhancement of the mixing between two neutrino flavours. For the parameters $\Delta m^2 \equiv m_2^2 - m_1^2$ and $\sin^2 2\vartheta$ ($m_1$ and $m_2$ are neutrino masses and $\vartheta$ is the vacuum mixing angle) a small mixing angle solution ($\Delta m^2 \simeq 5 \times 10^{-6}$ eV$^2$ and $\sin^2 2\vartheta \simeq 8 \times 10^{-3}$) and a large mixing angle solution ($\Delta m^2 \simeq 10^{-5}$ eV$^2$ and $\sin^2 2\vartheta \simeq 0.8$) were obtained. These solutions were found, however, from relatively low statistics data and under the assumption that the Standard Solar Model (SSM) [2] correctly predicts the neutrino fluxes from different reactions of the $pp$ and CNO cycles.

We will discuss here [3, 4, 5] the possibilities of a model independent treatment of the data of the future solar neutrino experiments (SNO [6], Super-Kamiokande (S-K) [7], ICARUS [8], etc.). In these experiments high energy neutrinos mostly from $^8$B decay will be detected. The distinctive new feature of the SNO experiment will be that in this experiment solar neutrinos will be detected through the simultaneous observation of the CC, NC and neutrino-electron elastic scattering (ES) processes

$$\nu_e + d \to e^- + p + p \qquad \text{(CC)} \qquad (1)$$
$$\nu + d \to \nu + p + n \qquad \text{(NC)} \qquad (2)$$
$$\nu + e^- \to \nu + e^- \qquad \text{(ES)} \qquad (3)$$

We will show here that in the general case of transitions of solar $\nu_e$'s into $\nu_\mu$ and/or $\nu_\tau$ the initial $^8$B neutrino flux can be determined independently on what is going on with neutrinos on their way from the sun to the earth. This will allow to determine the $\nu_e$ survival probability directly from the experimental data. We will show also that the data of the SNO and S-K experiments will allow to check in a model independent way whether there are sterile neutrinos in solar neutrino flux on the earth.

## 2. Transitions of solar $\nu_e$ into $\nu_\mu$ and/or $\nu_\tau$

The spectrum of the initial $^8$B neutrinos is given by the expression

$$\phi_{\nu_e}^{^8\mathrm{B}}(E) = \Phi_{\nu_e}^{^8\mathrm{B}} X(E) \qquad (4)$$

where $X(E)$ is a known function (the phase space factor of the decay $^8\mathrm{B} \to {^8\mathrm{Be}} + e^+ + \nu_e$) normalized by the condition $\left( \int X(E)\, \mathrm{d}E = 1 \right)$. The total flux $\Phi_{\nu_e}^{^8\mathrm{B}}$ is determined by number of $^8$B nuclei in the active zone of the sun and strongly depends on many parameters (temperature, cross-sections of different reactions and so on).

The flux of solar $\nu_e$'s on the earth as a function of neutrino energy $E$ will be determined through the observation of CC-induced events (1):

$$\phi_{\nu_e}(E) = n^{\mathrm{CC}}(E) / \sigma_{\nu_e d}^{\mathrm{CC}}(E) \qquad (5)$$

where $n^{\mathrm{CC}}(E)$ is the differential CC event rate and $\sigma_{\nu_e d}^{\mathrm{CC}}(E)$ is the total cross section of process (1). Furthermore, through the observation of CC and NC or CC and ES events the following ratios can be determined

$$R^{\mathrm{NC}} = 1 - \int \sigma_{\nu d}^{\mathrm{NC}}(E) \phi_{\nu_e}(E)\, \mathrm{d}E\, /\, N^{\mathrm{NC}}$$

$$R^{\mathrm{ES}} = 1 - \int \sigma_{\nu_e e}(E) \phi_{\nu_e}(E)\, \mathrm{d}E\, /\, N^{\mathrm{ES}}$$

where $N^{\mathrm{NC}}$ and $N^{\mathrm{ES}}$ are the total NC and ES event rates; $\sigma_{\nu d}^{\mathrm{NC}}(E)$ and $\sigma_{\nu_e e}(E)$ are the cross sections of the $\nu d \to \nu p n$ and $\nu_e e^- \to \nu_e e^-$ processes, respectively. If $R^{\mathrm{NC}} > 0$ ($R^{\mathrm{ES}} > 0$) it will be a direct proof that there are $\nu_\mu$ and/or $\nu_\tau$ in the solar neutrino flux on the earth.

The total flux of $^8$B neutrinos can be measured through the observation of NC events. In fact we have

$$\Phi_{\nu_e}^{^8\text{B}} = N^{\text{NC}}/X_{\nu d}^{\text{NC}} \qquad (6)$$

where $X_{\nu d}^{\text{NC}} \equiv \int \sigma_{\nu d}^{\text{NC}}(E)\, X(E)\, \mathrm{d}E = 4.1 \times 10^{-43}\,\text{cm}^2$. A comparison of the results of the measurement of the total flux with the SSM will be a severe test of the model.

For the $\nu_e$ survival probability we have

$$\text{P}_{\nu_e \to \nu_e}(E) = \frac{X_{\nu d}^{\text{NC}}}{\sigma_{\nu_e d}^{\text{CC}}(E)\, X(E)} \frac{n^{\text{CC}}(E)}{N^{\text{NC}}}$$

Thus the survival probability of solar $\nu_e$ as a function of neutrino energy $E$ can be determined directly from the data of the SNO experiment without any assumption about the initial $^8$B neutrino flux.

The total $^8$B neutrino flux can be determined also from the observation of CC and ES events. We have

$$\Phi_{\nu_e}^{^8\text{B}} = \Sigma^{\text{ES}}/X_{\nu_\mu e}^{\text{ES}} \qquad (7)$$

where

$$\Sigma^{\text{ES}} \equiv N^{\text{ES}} - \int \left(\sigma_{\nu_e e}(E) - \sigma_{\nu_\mu e}(E)\right) \phi_{\nu_e}(E)\, \mathrm{d}E$$

and $X_{\nu_\mu e}^{\text{ES}} \equiv \int \sigma_{\nu_\mu e}(E)\, X(E)\, \mathrm{d}E = 2.7 \times 10^{-45}\,\text{cm}^2$. From Eq.(6) and Eq.(7) it follows that the quantities $N^{\text{NC}}$ and $\Sigma^{\text{ES}}$ that will be measured in the SNO and S-K experiments are not independent and are connected by the relation $N^{\text{NC}} = r\, \Sigma^{\text{ES}}$, where $r \equiv X_{\nu d}^{\text{NC}}/X_{\nu_\mu e}^{\text{ES}} = 1.5 \times 10^2$. This relation can be violated only if there are transitions of solar $\nu_e$'s into sterile states.

### 3. Sterile Neutrinos

The problem of existence of sterile neutrinos is of principal importance for the theory of neutrino mixing (see ref.[9]). Future solar neutrino experiments could allow to reveal whether there are transitions of solar $\nu_e$ into sterile states [4, 5]. Using only the unitarity of the mixing matrix (conservation of probability) and excluding the initial flux, in the general case of active and sterile neutrinos instead of the relation $N^{\text{NC}} = r\, \Sigma^{\text{ES}}$ we have

$$N^{\text{NC}} - r\, \Sigma^{\text{ES}} = r \int \sigma_{\nu_\mu e}(E)\, \phi_{\nu_\text{S}}(E)\, \mathrm{d}E$$
$$- \int \sigma_{\nu d}^{\text{NC}}(E)\, \phi_{\nu_\text{S}}(E)\, \mathrm{d}E \qquad (8)$$

where $\phi_{\nu_\text{S}}(E)$ is the flux of sterile neutrinos on the earth. If it will occur that the left hand side of this equation, in which only measurable quantities enter, is found to be not equal to zero, it will mean that in the flux of solar neutrinos on the earth besides active neutrinos there are also sterile neutrinos. However, according to our model calculations the two terms in the right-hand side of Eq.(8) could cancel each other. In ref.[5] we proposed other relations which will allow to reveal the presence of sterile neutrinos. In general, for the average probability we have

$$\left\langle \sum_{\ell=e,\mu,\tau} \text{P}_{\nu_e \to \nu_\ell} \right\rangle_{\text{NC}} = \frac{N^{\text{NC}}}{X_{\nu d}^{\text{NC}}\, \Phi_{\nu_e}^{^8\text{B}}}$$

$$\left\langle \sum_{\ell=e,\mu,\tau} \text{P}_{\nu_e \to \nu_\ell} \right\rangle_{\text{ES}} = \frac{\Sigma^{\text{ES}}}{X_{\nu_\mu d}^{\text{ES}}\, \Phi_{\nu_e}^{^8\text{B}}}$$

where

$$\left\langle \sum_{\ell=e,\mu,\tau} \text{P}_{\nu_e \to \nu_\ell} \right\rangle_{\text{NC}}$$
$$\equiv \frac{\int \sigma_{\nu d}^{\text{NC}}(E)\, X(E) \sum_{\ell=e,\mu,\tau} \text{P}_{\nu_e \to \nu_\ell}(E)\, \mathrm{d}E}{X_{\nu d}^{\text{NC}}}$$

$$\left\langle \sum_{\ell=e,\mu,\tau} \text{P}_{\nu_e \to \nu_\ell} \right\rangle_{\text{ES}}$$
$$\equiv \frac{\int \sigma_{\nu_\mu e}(E)\, X(E) \sum_{\ell=e,\mu,\tau} \text{P}_{\nu_e \to \nu_\ell}(E)\, \mathrm{d}E}{X_{\nu_\mu e}^{\text{ES}}}$$

Furthermore, taking into account that $\text{P}_{\nu_e \to \nu_e}(E) = \phi_{\nu_e}(E)/\Phi_{\nu_e}^{^8\text{B}} X(E)$ and hence $\Phi_{\nu_e}^{^8\text{B}} \geq (\phi_{\nu_e}/X)_{\max}$, for the average probability of transition of solar $\nu_e$'s into all possible sterile states we obtain the following lower bounds:

$$\langle \text{P}_{\nu_e \to \nu_\text{S}} \rangle_{\text{NC}} \geq 1 - \mathcal{R}^{\text{NC}}$$
$$\langle \text{P}_{\nu_e \to \nu_\text{S}} \rangle_{\text{ES}} \geq 1 - \mathcal{R}^{\text{ES}}$$

where

$$\mathcal{R}^{\text{NC}} \equiv N^{\text{NC}}/X_{\nu d}^{\text{NC}}\, (\phi_{\nu_e}/X)_{\max}$$
$$\mathcal{R}^{\text{ES}} \equiv \Sigma^{\text{ES}}/X_{\nu_\mu e}^{\text{ES}}\, (\phi_{\nu_e}/X)_{\max}$$

are measurable quantities. If it will occur that $\mathcal{R}^{\text{NC}} < 1$ and/or $\mathcal{R}^{\text{ES}} < 1$ it will mean that there are transitions of solar $\nu_e$'s into sterile states. It is easy to see that the ratios $\mathcal{R}^{\mathrm{NC}}$ and $\mathcal{R}^{\mathrm{ES}}$ can be less than 1 only if the transition probability of solar $\nu_e$'s into all active states depends on neutrino energy $E$. Thus, if $\mathcal{R}^{\mathrm{NC}} < 1$ and/or $\mathcal{R}^{\mathrm{ES}} < 1$ it will mean not only that sterile neutrinos exist but also that the transition probability of solar $\nu_e$'s into sterile states depends on energy.

In S-K and other future solar neutrino experiments (ICARUS, etc.) a large number of solar neutrino induced ES events will be observed. From these data the differential ES event rate $n^{\mathrm{ES}}(E)$ will be determined and new possibilities for testing the existence of sterile neutrinos will emerge. In fact, a measurement of $n^{\mathrm{ES}}(E)$ and $n^{\mathrm{CC}}(E)$ will allow to determine the differential flux of all types of active neutrinos on the earth:

$$\sum_{\ell=e,\mu,\tau} \phi_{\nu_\ell}(E) = \frac{n^{\mathrm{ES}}(E)}{\sigma_{\nu_\mu e}(E)} + \left(1 - \frac{\sigma_{\nu_e e}(E)}{\sigma_{\nu_\mu e}(E)}\right) \frac{n^{\mathrm{CC}}(E)}{\sigma^{\mathrm{CC}}_{\nu_e d}(E)}$$

From the conservation of probability we obtain

$$\frac{1}{X(E)} \sum_{\ell=e,\mu,\tau} \phi_{\nu_\ell}(E) = \Phi^{^8\mathrm{B}}_{\nu_e} \left[1 - \mathrm{P}_{\nu_e \to \nu_{\mathrm{S}}}(E)\right]$$

If it will occur that the left-hand side of this equation, which contains only measurable quantities, depends on energy, then it will mean that there are sterile neutrinos in the flux of solar neutrinos on the earth. Furthermore, in the case under consideration we can obtain a lower bound for the transition probability of solar $\nu_e$'s into sterile states as a function of energy:

$$\mathrm{P}_{\nu_e \to \nu_{\mathrm{S}}}(E) \geq 1 - \frac{\sum_{\ell=e,\mu,\tau} \phi_{\nu_\ell}(E)}{X(E) \left(\sum_{\ell=e,\mu,\tau} \phi_{\nu_\ell}/X\right)_{\max}}$$

**4. Conclusions**

We have demonstrated that future solar neutrino experiments in which neutrino will be detected through the observation of *different reactions* have a great potential for a model independent investigation of the neutrino mixing problem. These experiments could allow also to *measure* the flux of initial $^8$B $\nu_e$'s and to test the SSM.